\documentclass[10pt,letterpaper]{article}
\usepackage{setspace}
\usepackage{amsmath}
\usepackage{graphicx}

\begin{document}

\title{Single-image separation measurements of two unresolved fluorophores}

\author{Shawn H. DeCenzo, Michael C. DeSantis,  and Y. M. Wang$^{\ast}$}

\date{Department of Physics, Washington University, St. Louis, MO 63130, USA}

\maketitle

\begin{table}
\centering
\small 
ymwang@wustl.edu
\end{table}

\pagebreak

\begin{abstract}
Measuring subdiffraction separations between single fluorescent particles is important for biological, nano-, and medical-technology studies.  Major challenges include (i) measuring changing molecular separations with high temporal resolution while (ii) using identical fluorescent labels.  Here we report a method that measures subdiffraction separations between two identical fluorophores by using a single image of milliseconds exposure time and a standard single-molecule fluorescent imaging setup.  The fluorophores do not need to be bleached and the separations can be measured down to 40 nm with nanometer precision.  The method is called single-molecule image deconvolution -- SMID, and in this article it measures the standard deviation (SD) of Gaussian-approximated combined fluorescent intensity profiles of the two subdiffraction-separated fluorophores.  This study enables measurements of (i) subdiffraction dimolecular separations using a single image, lifting the temporal resolution of seconds to milliseconds, while (ii) using identical fluorophores. The single-image nature of this dimer separation study makes it a single-image molecular analysis (SIMA) study.
\end{abstract}

\section{Introduction \label{introduction}}

For conventional far-field imaging, the Rayleigh criterion determines the diffraction limit seperation below which two identical fluorescent particles cannot be differentiated.  The diffraction-limit separation is $0.61\lambda$/N.A. $\approx 230$ nm for mean visible wavelength $\lambda$ of $\approx$ 550 nm and microscope objective's numerical aperture N.A. of $\approx$ 1.45 \cite{Born1999}.  In biological, nano-, and medical-technology studies, many inter- and intra-molecular separations are below the diffraction limit.  These separations include intra-molecular separations such as the separation between different regions of a macromolecule -- an actin filament or a chromosomal DNA molecule \cite{Wright1997}, and inter-molecular separations, \textit{e.g.}, between proteins and nano-particles \cite{Leong2003,Rotello2007}.  

Recent single-molecule fluorescence imaging methods circumvent the Rayleigh criterion and measure subdiffraction molecular separations with nanometer precision.  However, such precision is not achieved without tradeoffs.  Some methods bleach the fluorophores -- SHRImP \cite{Selvin2004}, NALNS \cite{Scherer2004}, PALM \cite{Hess2006}, and PAINT \cite{Hochstrasser2006}, and therefore only one separation of the studied molecules can be measured; some require photoswitchable or multicolor fluorescent labels -- Ref. \cite{Weiss2000}, PALM \cite{Hess2006}, SHREC \cite{Spudich2005}, and STORM \cite{Zhuang2007}, rendering the fluorophore and spectral selections restrictive.  In addition, these methods are based on centroid location measurements of each constituent molecule; therefore, the rate of separation measurements is bound by the total time required to locate all constituent molecules -- typically in seconds.  The timescale of seconds is long for many biological processes where molecular separations change faster than seconds.  Although the SHREC method allows simultaneous dimolecular separation measurements in the timescale of milliseconds, labeling heterogeneous fluorophores and overlapping two spectral distinct emission images pose additional challenges.

Here we report an alternative method to measure subdiffraction separations of two molecules labeled with identical fluorophores, all by using a single image of the two constituent molecules.  For simplicity, in the rest of this article we call the two separated fluorophores a dimer.  The dimers were simultaneously illuminated for milliseconds by standard laser excitation (intensity of order 1 kW/cm$^2$) and the emitted photons were collected by a standard single-molecule fluorescent imaging setup.  The collective intensity profile of the subdiffraction separated dimers were recorded and fit to a Gaussian function.  While the point spread function (PSF) centroids of the two constituent molecules cannot be resolved, the spread of the dimer intensity profile, measured as standard deviation (SD) of the dimer's Gaussian-approximated intensity profile (in this article we call the dimer intensity profile dimer PSF), increases with dimer separation.  In our method, the longitudinal SD (along the dimer direction) and the total detected photon count of the dimer image are the principle parameters needed for (1) differentiating dimers from monomers and (2) determining the separation of the dimer with known precision.  We have compared the experimental results with simulations and they agree well.

The reported SD study here analyzes a convolved image of two molecules and extracts the molecular separation information; the method is analogous to deconvolving a complex single-molecule image for molecular properties.  We call this method of using the spread of convolved single-molecule images to analyze molecular properties ``Single-molecule image deconvolution," or SMID.  Using SMID, dimer separations can be measured by using a single image of milliseconds exposure times, allowing improvement in temporal resolution from seconds to milliseconds.  The improvement is 10- to 1000-fold depending on the number of photons emitted by the dimer and the dimer separation.   This improved temporal resolution is attributed to the single-image nature of the SMID dimer separation measurements; here we name single-molecule studies that use single images of millisecond exposure times the ``Single-image molecular analysis" method, or SIMA.  This article applies SMID and SIMA to a simple convolution example of two molecules,  future SMID and SIMA analysis can be extended to higher order multimers of different geometrical arrangements.
\section{Results}
\subsection{Dimer PSF SD is an indicator of the subdiffraction separation} When two identical fluorophores are separated by less than the diffraction limit, the width of the dimer PSF varies with the dimer separation.  

\begin{figure}
\centering 
\includegraphics[width=3.5in]{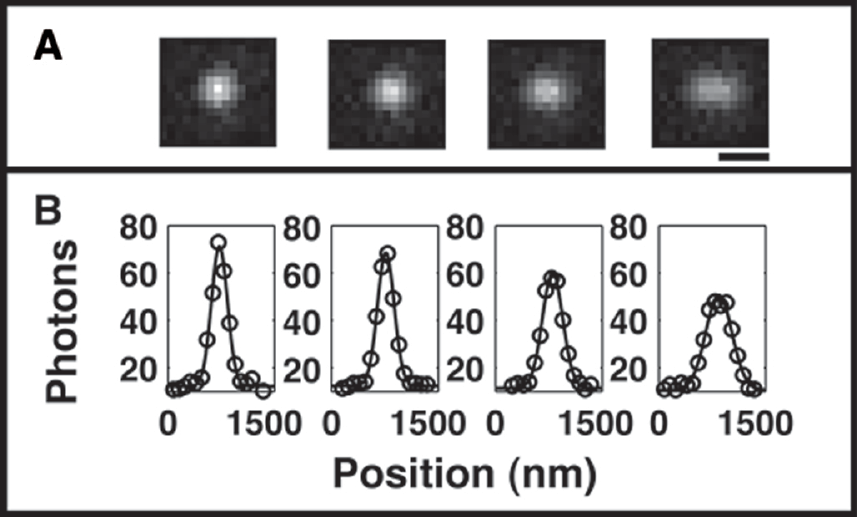}
	\caption{Dimers of different subdiffraction separations.  (A) From left to right, streptavidin-Cy3 dimers aligned in the horizontal direction separated by 0, 79, 158, and 237 nm.  From the appearance, it is not clear that the image contains two molecules until near the 237 nm separation.  The scale bar is 0.5 $\mu$m.  (B) 1D intensity profiles of the dimers (circles are data and lines are the Gaussian fits) showing that the dimer SDs increase with separation.}\label{Fig1} 
\end{figure}

Figure~\ref{Fig1}(A) shows an array of Streptavidin-Cy3 dimers with increasing horizontal subdiffraciton separations, $\delta$, of 0, 79, 158, and 237 nm.  The dimer images were constructed from two images of an experimental monomer movie: one of the two monomer images was displaced by 0, 1, 2, and 3 pixels (79 nm/pixel) horizontally, then the images were added.  The number of detected photons, $N$,  of each monomer image was 511 photons.  From the figures, it is clear that the width of the dimer images changes with separation.  At the 79 nm separation, the dimer is not clearly different from the constructed monomer (dimer at 0 nm separation);  at the 158 nm separation, the image becomes wider and it resembles a dimer more than a monomer; at the 237 nm separation, which is approximately equal to the diffraction limit of 230 nm, the image is clearly different from a monomer and the two monomer PSF peaks may start to be resolved.  

In Fig.~\ref{Fig1}(B) the dimer's increasing width with separation is quantified.  The dimer PSFs is unimodal as the sum of two Gaussian functions \cite{Behboodian1970} and fits well to a 1D Gaussian, where all transverse pixel intensity values at each longitudinal pixel were averaged.  The 1D $x$-axis SD values of the dimer increase with separation as 114.1, 119.7, 141.6, and 178.6 nm, respectively.

\subsection{Differentiating dimers from monomers using a single image}
As shown in Fig. \ref{Fig1}(A), for many subdiffraction separations, the dimer images are similar to monomer images in appearance.  In order to measure dimer separations, one must first determine whether an image is a monomer or a dimer.   Here we show that by using the PSF SD and photon count of a single image, dimers can be differentiated from monomers.  

For a monomer and a subdiffraction-separated dimer of the same photon count, the image is a dimer if the dimer PSF SD ($s_d$) exceeds the monomer PSF SD ($s_m$) by more than the sum of the dimer and monomer SD measurement errors, $\Delta{s_d}$ and $\Delta{s_m}$, respectively, as
\begin{equation}
s_{d} - s_{m} > \Delta s_{d} + \Delta s_{m}.
\label{dimer}
\end{equation}
Here the SD of a SD distribution, $\Delta s_{m}$ or $\Delta s_{d}$, is the error associated with a SD measurement of a single image.

\begin{figure*}
\centering
\includegraphics[width=5.3in]{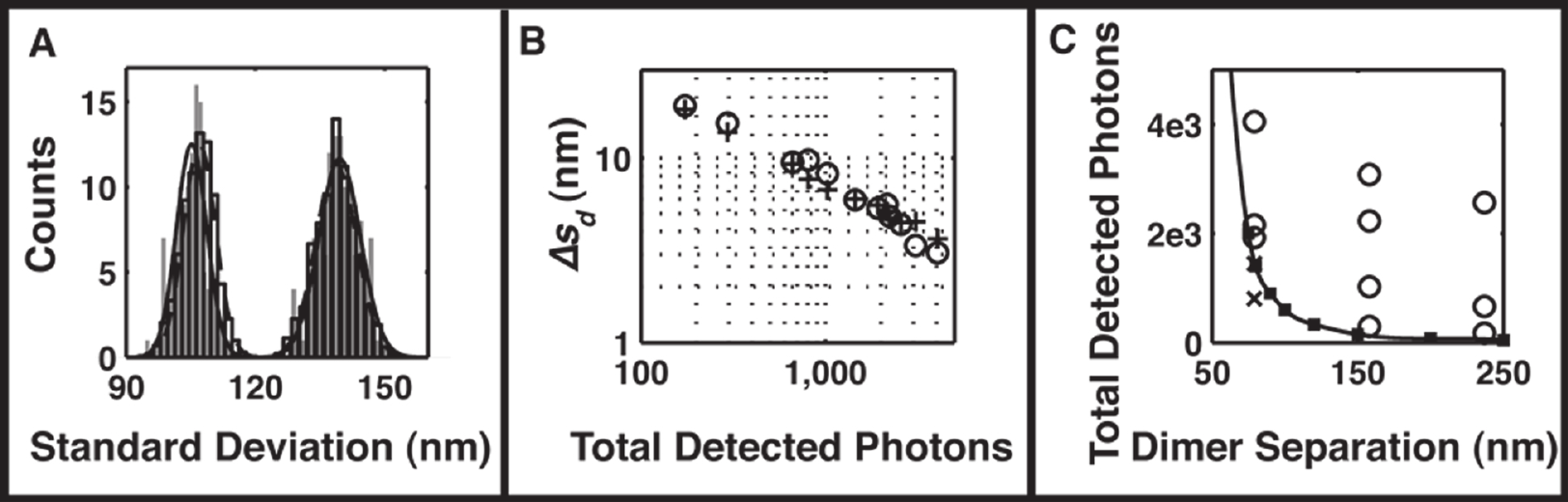}
	\caption{Distinguishing dimers from monomers.  (A) Experimental (grey) and simulated (empty bars) SD distributions of a constructed streptavidin-Cy3 monomer movie (left) and dimer movie (right) of separation 158 nm with the same photon count distribution of $N$ = 2200 $\pm$ 156 photons (mean $\pm$ SD).  The Gaussian fits to the experimental (solid line) and simulation data (dashed lines) are shown.  The simulations were for 1000 runs and the histograms were scaled to be comparable to the experimental data.  (B) Comparing experimental (circles)   and simulation (crosses) SD measurement error $\Delta{s_d}$ of a 158 nm separated dimer at different detected photon counts.  (C) Threshold photon count line (spline line) for distinguishing dimers from monomers at different separations.  The points on the line are simulation data; circles are experimental data above the line, indicating differentiable dimers, and crosses are experimental data below the line, indicating monomers or undifferentiable dimers.}\label{Fig2}
\end{figure*}

Figure \ref{Fig2}(A) compares the experimental and simulation SD distributions of a constructed monomer (0 nm separated dimer) and a 158 nm separated dimer.  The two experimental distributions were constructed from one monomer movie, which has 236 valid images with photon count of 1,100 $\pm$ 124 (mean $\pm$ SD).  After construction, the mean dimer movie photon count was 2,200, and the number of images was 118.  The simulations used the experimental monomer $N$ and $s_m$ distributions and the experimental 0 nm separation dimer background photon distributions.  The simulations were run for 1000 iterations, and in Fig.~\ref{Fig2}(A) the counts were scaled to have the same amplitude as the experimental distributions for comparison.  The experimental distributions agree with the simulations.

For the experimental distributions, the Gaussian fits yield the means to be $s_m$ = 105.2 nm and $s_d$ = 139.5 nm, and SDs to be $\Delta s_{m}$ = 3.5 nm and $\Delta s_{d}$ = 3.8 nm, respectively.  For the simulated distributions, $s_{m}$ = 106.7 nm and $s_d$ = 138.6 nm, and $\Delta s_{m}$ = 3.8 nm and $\Delta s_{d}$ =  4.8 nm, respectively.  From the information of Fig. \ref{Fig2}(A), we can conclude that for an image with a photon count above 2200 and $s_d$ above $s_{m} + \Delta s_{d} + \Delta s_{m} = $ 106.7 nm + 3.8 nm + 4.8 nm = 115.3 nm (the simulation results), it is a dimer.  Our 158 nm separated dimer has a mean $s_d$ of 139.5 nm $>$ 115.3 nm, therefore the 158 nm separation images are dimers.  

The dimer SD measurement error, $\Delta s_{d}$, is affected by the total number of detected photons in the image.  Figure \ref{Fig2}(B) shows the experimental and simulated $\Delta s_{d}$ dependence on photon counts for the 158 nm separation.  $\Delta s_{d}$ decreases with increasing photon counts, and the experimental results agree well with simulations (Fig. \ref{Fig2}).  There are a total of 12 monomer movies of different photon counts used for the figure, and all experimental data in this article were selected from these data.  These monomer data are the same data as that in our previous article (\cite{Wang2010,Wang2009}).     

In order to differentiate dimers from monomers, Fig. \ref{Fig2}(C) shows the threshold photon count vs. separation diagram.  At a certain separation, if an image's photon counts is above the threshold value on the line, then it is a dimer; below the line, it is not differentiable.  The line was constructed by dimer simulations of different separations and photon counts.  At a particular separation, when the condition of $s_{d} - s_{m} = \Delta s_{d} + \Delta s_{m}$ was met, the photon count was the threshold photon count for the separation.  The 12 experimental data were used to construct dimers each at a random separation.  By using Fig. \ref{Fig2}(A) distributions and Eq.~\ref{dimer} on each monomer data, the differentiable dimers were plotted as circles, and undifferentiable dimers were plotted as crosses.  The experimental data agree with the diagram that all differentiable dimers were above the line and undifferentiable dimers were below the line.

\subsection{Dimer separation measurements}
\begin{figure*}
\centering
\includegraphics[width=3.5in]{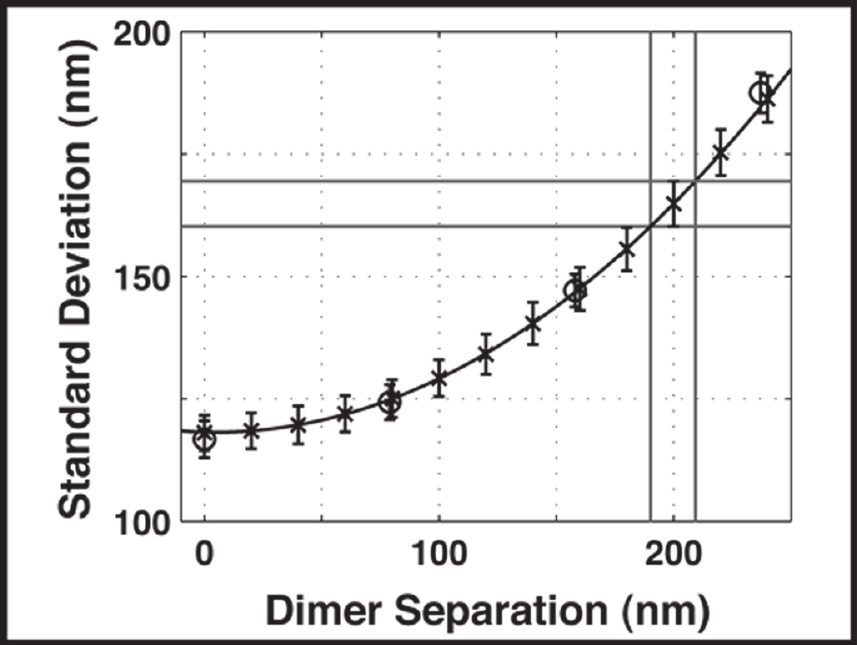}
	\caption{Dimer SD vs. separation diagram.  The crosses are the mean SD values for simulations of 1000 dimer images, and the circles are experimental data at the separations of 0, 79, 158, and 237 nm.  The error bars are the SDs of the SD distributions.  A fit to the simulation results is the solid line.  The horizontal grey lines off the extrema of the error bar of the 200 nm separation data meet the diagram, measuring the dimer separation measurement errors by the vertical grey lines.}\label{Fig3} 
\end{figure*}

After determining that an image is a dimer, the next goal is to determine the dimer separation.  Figure \ref{Fig3} shows the experimental (circle) and simulation (crosses) dimer PSF SD vs. separation diagram.  The SD data are means of dimer SD distributions, and the error bars are the SDs of the dimer SD distributions.  The experimental results at four different separations were constructed from a single monomer movie.  The mean experimental dimer photon count was 3050.  Both experimental $s_d$ and $\Delta{s_d}$ results agree with the simulation results.  For simplicity, the solid line is a fit to the simulation results as $s_d = 0.001\delta^2-0.01\delta+118$, and is independent of $N$.  This diagram allows for direct translation of dimer PSF SD to separation with known uncertainty determined by the image's photon count, as described below.

\subsection{Dimer separation measurement error} 
\begin{figure*}
\centering
\includegraphics[width=3.5in]{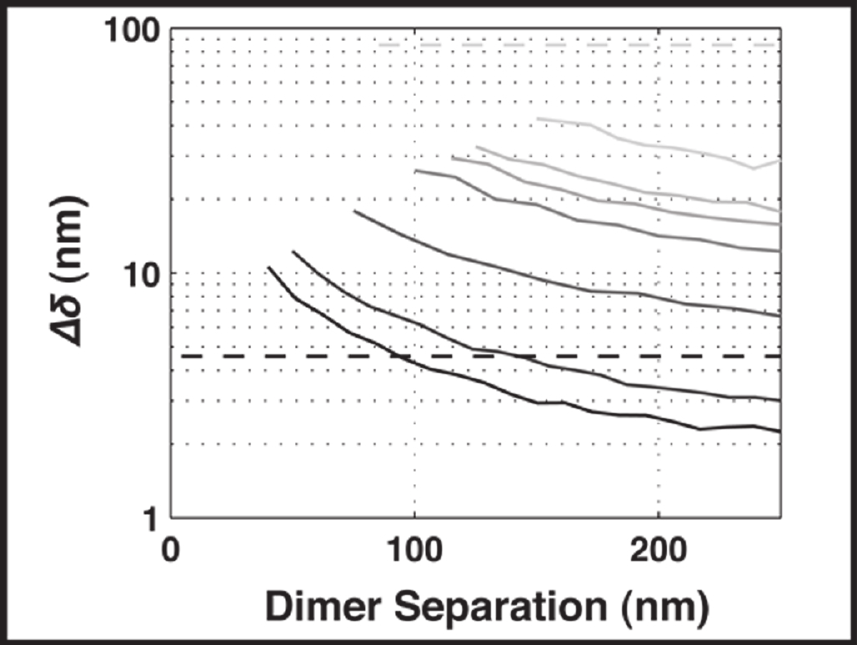}
	\caption{Dimer separation measurement error, $\Delta\delta$ vs. separation at different photon counts of 150, 300, 400, 600, 2000, 10000, and 20,000 (slanted lines from light to dark).  The low separation termination points of the lines are the separations below which dimers are not differentiable from monomers.   The horizontal dashed lines are the dimer separation measurement error by using the centroid measurement method for dimer photon counts of 150 (top line) and 20,000 (bottom line).}\label{Fig4} 
\end{figure*}

Each dimer separation measurement should have an associated precision.  Figure~\ref{Fig4} shows the dimer separation uncertainty $\Delta{\delta}$ vs. separation $\delta$ for different numbers of detected photons from 150 to 20,000.  The separation uncertainty $\Delta{\delta}$ was obtained from the dimer SD measurement error bars, as the ones shown in Fig.~\ref{Fig3} for the 3050 photon dimers.  The ends of the dimer error bars were extrapolated horizontally (horizontal grey lines) to meet with the $s_d$ vs. $\delta$ line.  The separation values at the two cross points on the $s_d$ vs. $\delta$ line (vertical grey lines) marked the dimer's two separation measurement deviations.  The differences between the cross point separation values and the mean separation are the upper and lower errors for this separation measurement.  The average of the two errors was the dimer separation measurement error $\Delta{\delta}$ for the specific separation and photon count.  A different $s_d$ vs. $\delta$ simulation was run for each different photon count.  

In Fig.~\ref{Fig4}, there are four properties to the lines:  (1) On the left, each line terminates at the dimer-monomer differentiation threshold for the photon count [Eq.~\ref{dimer} and Fig.~\ref{Fig2}(C)].  (2) The line values decrease with increasing separation.  This is because that although the dimer SD error is approximately independent of the separation, the $s_d$ vs. $\delta$ line flattens out at small separations, identical SD error translates into larger separation measurement error at low separations.  As a result, the $\Delta\delta$ line decreases with $\delta$. (3) As the photon count increases, $\Delta\delta$ decreases to the nanometer range.  For 20,000 photons, $\Delta\delta$ is $\approx$ 2.3 nm at the 250 nm separation and 10 nm at the 40 nm separation.  For 150 photons, $\Delta\delta$ is 42 nm at the 250 nm separation and 29 nm at the 150 nm separation.  (4) Dimer separation measurement errors by using the centroid method are shown as the two horizontal dashed lines for 150 and 20,000 photons.  Comparing to the centroid method, the precision of the SD measurement method is comparable at high photon counts and higher at low photon counts.

The upper bound of dimer separation measurement errors by using centroid measurements is the sum of the two centroid measurement errors, one for each monomer at half of the detected photon count of the dimer.   The centroid measurement errors were calculated using the analytical expression in Ref. \cite{Wang2010,Wang2009} for the corresponding photon counts and background information, plus 40$\%$ increase to accommodate for the discrepancy between experimental and analytical results \cite{Wang2010,Wang2009,Webb2002}.  For 150 photons (upper dashed line), the sum of the two centroid measurement errors is (2 $\times$ 30.5 nm)$\times$ 1.4 = 85.3 nm.  For 20,000 photons (lower dashed line), the sum of the two centroid errors is $\approx$  (2 $\times$ 1.6 nm) $\times$ 1.4 = 4.5 nm.  The low separation end points of the centroid error lines are the magnitude of the centroid dimer separation measurement errors, which are 85.3 nm for 150 photons and 4.5 nm for 20,000 photons (the values agree with another estimation in Ref. \cite{Ober2006}).  

\subsection{Temporal resolution diagram}
 The single-image nature of the SD measurement method has determined that the temporal resolution of the dimer separation measurements is determined by the exposure time (or the camera's frame imaging speed, if slower).  The minimum dimer exposure time is determined by the minimum number of detected photons required to differentiate dimers from monomers, as indicated in Fig.~\ref{Fig2}(C).  
 
 \begin{figure*}
\centering
\includegraphics[width=3.5in]{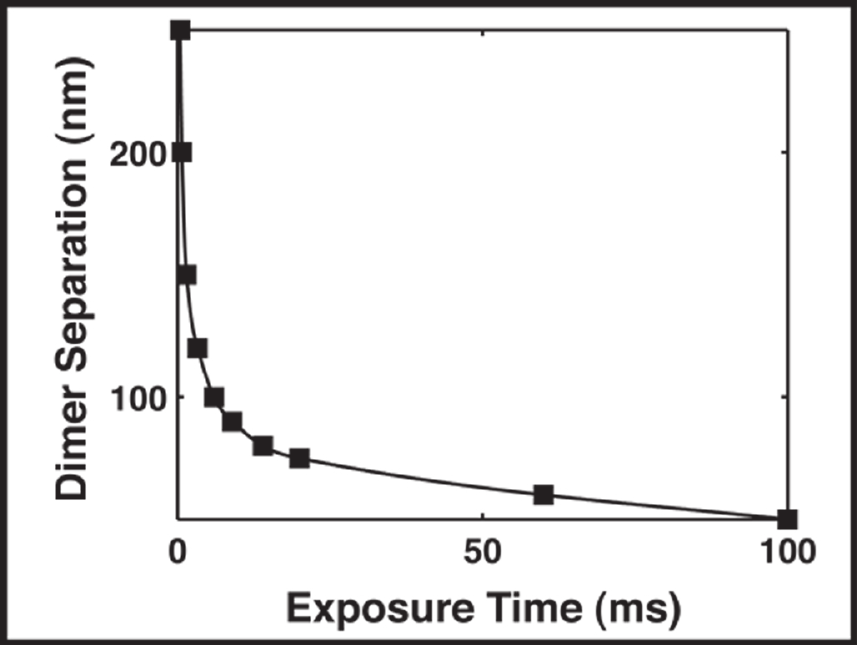}
	\caption{SMID dimer separation measurement temporal resolution diagram.  The line marks the threshold exposure time (threshold photon count) to differentiate dimers from monomers.  The data points were converted from points in Fig.~\ref{Fig2}(C) using the conversion factor of 100 detected photons per 1 ms of exposure time, and the line is the splined connection of the data.  The SMID timescales for all attainable subdiffraction dimer separation measurements are milliseconds.}\label{Fig5} 
\end{figure*}
 
Figure \ref{Fig5} shows the temporal resolution vs. dimer separation diagram.   In constructing the diagram, the required photons in Fig.~\ref{Fig2}(C) were converted into exposure times, assuming that (1) a typical laser intensity of 300 W/cm$^2$ is used, (2) the detected dimer photon count is 100 photons per 1 ms exposure time on average for our system, and (3) the number of dimer emitted photons increases linearly with the exposure time.  For the 20 ms exposure time, 2,000 photons will be detected and the dimer separation can be measured down to 75 nm with 18 nm precision.  For the 100 ms exposure time, 10,000 photons will be detected and the dimer separation can be measured down to 50 nm with 12 nm precision.  Note that all the exposure times are in the milliseconds range.  
\section{Discussion} \subsection{Advantages and limitations of SMID for subdiffraction dimer separation measurements} There are advantages and limitations to using SMID for subdiffraction dimer separation measurements.  The main advantages are that the method uses (1) a standard single-molecule imaging system and (2) only milliseconds of imaging time.  The main limitations are that (i) the minimum measurable dimer separation only reaches 40 nm for a reasonable 200 ms exposure time.  More photons are needed to measure smaller separations; although, with increasing exposure time the temporal resolution will be compromised; (ii) typical experimental 100 to 2000 photon counts per dimer image restricts the most useful range of measurable dimer separations to be between 70 nm, the typical pixel size, and 250 nm, the Rayleigh criteria.  

\subsection{Comparing spatial resolution with the centroid measurement method}  In measuring subdiffraction dimer separations using the centroid method (SHRImP, SHREC, PALM, and NALNS), two centroid measurements are performed (one for each monomer, or one for the dimer and the other for the second monomer after the first monomer bleaches).  Both methods reaches nanometer resolution in measuring dimer separations.  Comparing the spatial resolution of the two methods by using Fig.~\ref{Fig4}, there are following differences: (1) the precisions to dimer separation measurements is higher for the SD method at low photon counts and comparable at high photon counts; (2) at a fixed photon count, the centroid separation measurement error is constant and reaches a lower measurable separation minimum, while the SD measurement error increases with decreasing separation and terminates at a higher measurable minimum separation.  If temporal resolution is not a concern, future dimer subdiffraction measurements can use Fig.~\ref{Fig4} for choosing the appropriate method for optimal spatial resolution.

\subsection{Dimer orientation effects} In this article,  the two fluorophores are identical and both are located in the imaging plane.  Although the number of photon emitted per monomer differs from image to image due to stochastic fluctuation \cite{Wang2005}, the emitted photon distributions of the two monomers are the same.  In experiments, this is frequently not the case and most dimers are randomly oriented in 3D, rather than both being fixed at the imaging surface.  This can (1) cause difference in $s_m$ between the two molecules due to defocusing \cite{Florin2003}; (2) for total internal reflection fluorescence (TIRF) imaging microscopy studies, the detected photon counts of the defocused monomers will be less than that of the focused monomer due to the decaying evanescent light intensity \cite{Wang2006}, and as a result, (3) the dimer SD vs. separation relation in Fig.~\ref{Fig3} will alter.  Future separation measurements of different dimer orientations should include appropriate orientation adjustments in data analysis.   

\section{Conclusion}
In this article we report a simple method to measure subdiffraction separations between two identical fluorophores -- the SMID method -- that measures standard deviation of dimer intensity profiles.  This method is advantageous over existing methods by its applicability to any fluorophores and the fast dynamic rate in determining dimer separations by using only a single image of milliseconds exposure times by using SIMA.  Although the separation measurements only have a lower limit of 40 nm for 20,000 photons, it will meet the demand of many biological and nano-science applications that require dynamic separation measurements in the range of order 70 nm to 250 nm.   

\section{Methods}
\subsection{Sample preparation and imaging}
 Streptavidin-Cy3 powder (0.04 nM, SA1010, Invitrogen, Carlsbad, CA) was dissolved in 0.5X TBE buffer (45 mM Tris, 45 mM Boric Acid, 1 mM EDTA, pH 7.0) to make the protein solution.  Coverslips and fused-silica chips were cleaned using oxygen plasma.  Streptavidin-Cy3 molecules were immobilized on fused-silica surfaces by sandwiching 6 $\mu$l of the protein solution between the surface and a coverslip.  The coverslip edges were then sealed with nail polish.  

Single molecule imaging was performed using a Nikon Eclipse TE2000-S inverted microscope (Nikon, Melville, NY) in combination with a Nikon 100X objective (Nikon, 1.49 $N.A.$, oil immersion).  Samples were excited by prism-type TIRF microscopy with a linearly polarized 532 nm laser line (I70C-SPECTRUM Argon/Krypton laser, Coherent Inc., Santa Clara, CA) focused to a 40 $\mu$m $\times$ 20 $\mu$m region.  The laser excitation was pulsed with illumination intervals between 1 ms and 500 ms and excitation intensities between 0.3 kW/cm$^2$ and 2.6 kW/cm$^2$.  By combining laser intensity and pulsing interval variations we obtained 50 to 3000 detected photons per monomer PSF.  Images were captured by an iXon back-illuminated electron multiplying charge coupled device (EMCCD) camera (DV897ECS-BV, Andor Technology, Belfast, Northern Ireland).  An additional 2X expansion lens was placed before the EMCCD, producing a pixel size of 79 nm.  The excitation filter was 530 nm/10 nm and the emission filter was 580 nm/60 nm. 
\subsection{Data acquisition and selection}
Typical movies were obtained by synchronizing the onset of camera exposure with laser illumination for different intervals.  The gain levels of the camera were adjusted such that none of the pixels of a PSF reached the saturation level of the camera.  When single-molecule movies were obtained, streptavidin-Cy3 monomers were first selected in \textsc{ImageJ} (NIH, Bethesda, MD) by examining the fluorescence time traces of the molecules for a single bleaching step \cite{Wang2005}.  For a selected monomer, the camera intensity values for 25 $\times$ 25 pixels centered at the molecule were recorded.  One of two 25 $\times$ 25 pixels monomer boxes were shifted and added to the other box to create a dimer.  For the summed dimer images, the center 15 $\times$ 15 pixels were used for 2D Gaussian fitting of the 0 nm separated dimer, and the intensities from the peripheral pixels of the 0 nm dimer box were used for experimental dimer background analysis.  The center 20 $\times$ 20 pixels intensities were used for analysis of other dimer separations.  By selecting the center pixels for analysis, the non-overlapping background pixels were avoided.

Before analysis, the camera intensity count of each pixel in an image was converted into photon counts by using the camera-count to photon-count conversion factor calibrated in our previous article \cite{Wang2010,Wang2009}.  Then, the number of detected photons, and the $x$-axis and $y$-axis dimer PSF SD were obtained.  The number of detected photons was obtained by subtracting the total photon count of the image by the total photon count of the background; the two SD values were parameters of a 2D Gaussian fit to the intensity profile of the image using equation
\begin{equation}
f(x,y)=f_0  \exp{\left[-\frac{(x-x_0)^2}{2 s^2_x} - \frac{(y-y_0)^2}{2 s^2_y}\right]}+ \langle b \rangle,
\label{Gaussian}
\end{equation}
here $f_0$ is the multiplication factor, $s_x$ and $s_y$ are SD values in $x$ and $y$ directions, respectively, $x_0$ and $y_0$ are the centroid of the molecule, and $\langle{b}\rangle$ is the mean background offset in photon counts.  

The selected streptavidin-Cy3 monomers were further characterized to satisfy the following conditions used for dimer SD analysis.  (1) No stage drift detected by using centroid \textrm{vs} time measurements.  Stage drift introduces additional blur to each single-molecule PSF and thus affects the measured SD values.  (2) A minimum of 75 valid PSF images, each with a photon count, $N$, that fluctuated less than 20$\%$ from the experimental mean $\langle{N}\rangle$, of the monomer.  The PSF $N$ count restriction is necessary for precise SD error analysis at $N$ by using a statistically sufficient number of PSFs with consistent $N$.  (3) PSFs with signal-to-noise ratios ($I_0/\sqrt{I_0+\sigma_b^2}$) larger than 2.5, where $I_0$ is the peak PSF photon count (total photon count minus $\langle{b}\rangle$) and $\sigma_b^2$ is the background variance in photons.  (4) Mean monomer SD values, $\langle{s_{mx}}\rangle$ and $\langle{s_{my}}\rangle$, obtained by Gaussian fitting of the $s_{mx}$ and $s_{my}$ distributions of all valid images did not differ by more than 10 nm, or $\pm$ 5$\%$ of the mean SD value to minimize polarization effects of Cy3.  (5) The mean SD values $\langle{s_{mx}}\rangle$ and $\langle{s_{my}}\rangle$ were between 95 nm and 135 nm to minimize defocusing effects.  Images from the 12 monomer movies used in this article satisfy the described restrictions.

\subsection{Creating experimental dimer movies from monomer movies}
Experimental dimer images were constructed by adding all consecutive nondegenerate two images of a monomer movie, with one of the two images shifted 0 - 3 pixels in the $x$ direction.  The final dimer images were reboxed to the center 15 $\times$ 15 pixels for the 0 nm separation images and 20 $\times$ 20 pixels for other separations.  All experimental data in Fig.~\ref{Fig2} were constructed from the 12 monomer data movies with a selected separation.   For the experimental data in Fig.~\ref{Fig3}, one monomer movie was used for all different separations. 

\subsection{Dimer simulations}
To simulate dimers, we first generated monomers.  Single-fluorescent-molecule PSFs were generated using the Gaussian random number generator in \textsc{MATLAB}.  For simulations that later compare with experimental data (Figs. \ref{Fig2}(A), \ref{Fig2}(B), and \ref{Fig3}), the simulated monomer's SD without the pixelation effect, $s_{m0x,m0y}$, was determined by the experimental means $\langle{s_{mx,my}}\rangle$ after subtracting for the pixelation effect (Eq. 15 in Refs. \cite{Wang2010} and \cite{Wang2009}).  The finite bandwidth of the emission filter was also taken into consideration by simulating each photon as being drawn from a PSF whose width is varied according to a Gaussian distribution centered about $s_{m0x,m0y}$ (with SD of 2 nm).  The experimental $N$ distribution and the restriction that only photon count that fluctuated less than 20$\%$ from the mean $N$ were used.  For simulations that do not compare with experimental data (Figs. \ref{Fig2}(C) and \ref{Fig4}), the simulated monomer's SD was $s_{m0x,m0y}$ = 110 nm, the standard deviation in photon count was 10$\%$ of the mean photon count and again only randomly generated $N$ that stayed within 20$\%$ of the mean was used for images.  The generated photons of each PSF were binned into 20 $\times$ 20 pixels with a pixel size of 79 nm.  

Using the simulated monomer movie, two nondegenerate consecutive images were shifted and then sum to create dimers.   After construction of dimer intensity profile in photons, each photon in a pixel was converted into camera count using Eq. 6 in Refs. \cite{Wang2010} and \cite{Wang2009} with a conversion factor $M$ of one.  After the dimer PSF construction in camera counts, background photon distributions were added to the image.  For Figs. \ref{Fig2}(A), \ref{Fig2}(B), and \ref{Fig3}, random background photons at each pixel were generated using the corresponding experimental dimer background distribution functions obtained by using the 0 nm separation dimer.  For Figs. \ref{Fig2}(C) and \ref{Fig4}, the background photon distributions had the mean values of the experimental 0 nm dimer separation of all 12 monomer data: the mean dimer background photon counts was 1.8 photons, and the mean dimer background standard deviation was 1.7 photons.  From each final dimer image, the center 20 $\times$ 20 pixels were used for 2D Gaussian analysis. For each simulated dimer datum, 1000 iterations were performed.

\section*{Acknowledgements}
Michael C. DeSantis is supported by a National Institutes of Health predoctoral fellowship awarded under 5T90 DA022871.  Shawn H. DeCenzo is supported by an US Dept of Education
``Graduate Assistance in Areas of National Need" award under P200A090267.


\begin{thebibliography}{20}

\bibitem{Born1999} M. Born and E. Wolf, \textit{Principles of Optics} (Cambridge University Press, Cambridge, UK, 1999).

\bibitem{Wright1997} G. S. Gordon, D. Sitnikov, C. D. Webb, A. Teleman, A. Straight, R. Losick, A. W. Murray, and A. Wright, ``Chromosome and Low Copy Plasmid Segregation in \textsc{E}. coli: Visual Evidence for Distinct Mechnisms,'' Cell {\bf 90} 1113-1121 (1997).

\bibitem{Leong2003} A. K. Salem, P. C. Searson, and K. W. Leong, ``Multifunctional nanorods for gene delivery,'' Nature {\bf 2}, 668-671 (2003).

\bibitem{Rotello2007} G. Han, P. Ghosh, M. De, and V. M. Rotello, ``Drug and Gene delivery using gold nanoparticles,'' Nanobiotechnolgoy {\bf 3}, 40-45 (2007).

\bibitem{Selvin2004} A. Yildiz, M. Tomishige, R. D. Vale, and P. R. Selvin, ``Kinesin walks hand-over-hand,'' Science {\bf 303}, 676-678 (2004).

\bibitem{Scherer2004} X. Qu, D. Wu, L. Mets, and N. F. Scherer, ``Nanometer-localized multiple single-molecule fluorescence microscopy,'' Proc. Natl. Acad. Sci. USA {\bf 101}, 11298-11303 (2004).

\bibitem {Hess2006} E. Betzig, G. H. Patterson, R. Sougrat, O. W. Lindwasser, S. Olenych, J. S. Bonifacino, M. W. Davidson, J. L. Schwatz, and H. F. Hess, ``Imaging Intracellular Fluorescent Proteins at Nanometer Resolution,'' Science {\bf 313}, 1642-1645 (2006). 

\bibitem{Hochstrasser2006} A. Sharonov and R. M. Hochstrasser, ``Wide-field subdiffraction imaging by accumulated binding of diffusing probes,'' Proc. Natl. Acad. Sci. USA {\bf 103}, 18911-18916 (2006).

\bibitem{Weiss2000} T. d. Lacoste, X. Michalet, F. Pinaud, D. S. Chemla, A. P. Alivisatos, and S. Weiss, ``Ultrahigh-resolution multicolor colocalization of single fluorescent probes,'' Pro. Natl. Acad. Sci. USA {\bf 97}, 9461-9466 (2000).

\bibitem{Spudich2005} L. S. Churchman, Z. $\ddot{O}$kten, R. S. Rock, J. F. Dawson, and J. A. Spudich, ``Single molecule high-resolution colocalization of Cy3 and Cy5 attached to macromolecules measures intramolecular distances through time,'' Pro. Natl. Acad. Sci. USA {\bf 105}, 1419-1423 (2005).

\bibitem{Zhuang2007} M. Bates, B. Huang, G. T. Dempsey, X. Zhuang, ``Multicolor Super-Resolution Imaging with Photo-Switchable Fluorescent Probes,'' Science {\bf 317} 1749-1753 (2007).

\bibitem{Behboodian1970} J. Behboodian, ``On the modes of a mixture of two normal distributions,'' Technometrics {\bf 12} 131-139 (1970).

\bibitem{Wang2010} M. C. DeSantis, S. H. DeCenzo, Y. M. Wang, ``Precision analysis for standard deviation measurements of single-fluorescent molecule images,'' Opt. Express doc. ID 123194 (posted 15 March 2010, in press).

\bibitem{Webb2002} R. E. Thompson, D. R. Larson, and W. W. Webb,``Precise nanometer localization analysis for individual fluorescent probes,'' Biophys. J. {\bf 82}, 2775-2783 (2002).

\bibitem{Ober2006}S. Ram, E. S. Ward, R. J. Ober, ``Beyond Rayleigh's criterion: A resolution measure with application to single-colecule microscopy,'' Proc. Natl. Acad. Sci. USA {\bf 103}, 4457-4462 (2006).

\bibitem{Wang2005} Y. M. Wang, J. Tegenfeldt, W. Reisner, R. Riehn, X.-J. Guan, L. Guo, I. Golding, E. C. Cox, J. Sturm, and R. H. Austin, ``Single-molecule studies of repressor-\textsc{DNA} interactions show long-range interactions,'' Proc. Natl. Acad. Sci. USA {\bf 102}, 9796-9801 (2005).

\bibitem{Florin2003} M. Speidel, A. Jonas, and E.-L. Florin, ``Three-dimensional tracking of fluorescent nanoparticles with subnanometer precision by use of off-focus imaging,'' Opt. Lett. {\bf 28} (2003).

\bibitem{Wang2006} Y. M. Wang, R. H. Austin, and E. C. Cox, ``Single molecule measurements of repressor protein 1\textsc{D} diffusion on \textsc{DNA},'' Phys. Rev. Lett. {\bf 97}, 048302 (2006).

\end{thebibliography}
\end{document}